\documentstyle[12pt,mont98_desy,amssymb,epsfig]{article}
\begin{document}

\title{\Large \bf {Radiative Polarization, Computer Algorithms and Spin 
Matching in Electron Storage Rings}
\footnote{Contributions to the Handbook of Accelerator Physics and 
Engineering, Eds. A.W. Chao and M. Tigner, 1st edition, 3rd printing, World Scientific, 
2006.
%Also as DESY report 99-095.
} 
}

\author{  \large{D.P.~Barber and G. ~Ripken {\footnote { ~~$\dagger$ December 2004.}} }}

\address{Deutsches Elektronen--Synchrotron, DESY, \\
 22603 Hamburg, Germany. \\( E-mail: mpybar@mail.desy.de)}

\maketitle
\abstracts{We present a set of notes, meant for quick reference, on 
radiative spin polarization, computer algorithms and spin matching in
       electron storage rings.}

\section*{\large \bf{2.7.7.~Radiative Polarization in Electron Storage Rings}}

\paragraph{The Sokolov-Ternov effect \protect{\cite{st64}}}
      Relativistic electrons in a storage ring emit synchrotron radiation
(Sec.3.1 in \cite{handbooka}). A very small fraction of the
radiated photons cause spin
flip. For electron spins aligned along a uniform magnetic
field, the $ \uparrow\downarrow $ and $\downarrow\uparrow$ flip rates
differ and this leads to a build-up of spin polarization
antiparallel to the field. Positrons become polarized parallel to the field.
The transition rates for electrons are
\begin{eqnarray}
      W_{\uparrow\downarrow}
       &=&
      \frac{5\sqrt{3}}{16}
      \frac{r_{\rm e} \gamma^{5}\hbar}
           {m_{\rm e}|\rho|^{3}}
      \left(1+
      \frac{8}{5\sqrt{3}}
          \right)
 \nonumber \\
 \nonumber \\
      W_{\downarrow\uparrow}
       &=&
      \frac{5\sqrt{3}}{16}
      \frac{r_{\rm e} \gamma^{5}\hbar}
           {m_{\rm e}|\rho|^{3}}
      \left(1-
      \frac{8}{5\sqrt{3}}
    \right)
\label{SokolovTernov}
\end{eqnarray}
For positrons, interchange plus and minus signs here and elsewhere
\footnote{$\rho$ is the radius of curvature of the orbit, $r_{\rm e}$ is the classical electron radius and the other symbols
have their usual meanings.}.

The equilibrium polarization in a uniform magnetic
field is independent of $\gamma$,
\begin{eqnarray}
      P_{\rm st}
         &=&
      \frac{
            W_{\uparrow\downarrow}
                    -
            W_{\downarrow\uparrow}
                                    }
           {
            W_{\uparrow\downarrow}
                    +
            W_{\downarrow\uparrow}
                                    }
      \ =\
      \frac{8}{5\sqrt{3}}
     \ =\
    0.9238
\end{eqnarray}
For a beam with zero initial polarization,
the time dependence for build-up to equilibrium is
\begin{eqnarray}
    P(t)
       &=&
    P_{\rm st}
    \left[
          1-\exp{(-t/\tau_{0})}
              \right]
\end{eqnarray}
where  the build-up rate is
\begin{eqnarray}
    \tau_{0}^{-1}
      \ =\
      \frac{5\sqrt{3}}{8}
      \frac{r_{\rm e} \gamma^{5}\hbar}
           {m_{\rm e}|\rho|^{3}}
\label{eq:ST}
\end{eqnarray}
$\tau_0$ depends strongly on $\gamma$ and $\rho$
but is typically minutes or hours.
In a flat ring in which all bending magnets have the same $\rho$
just average Eq.(\ref{eq:ST}) over the circumference $C$:
\begin{equation}
     \tau_{0}^{-1} [{\rm s}^{-1}]
       \ \approx\
      \frac{2\pi}{99}
      \frac{E[{\rm GeV}]^{5}}
           {C[{\rm m}]\rho [{\rm m}]^{2}}
\end{equation}

\vspace{5mm}
\paragraph{The Baier-Katkov flip rate}
For electron spins initially aligned along an arbitrary
                                      unit vector $\hat\xi$ the
generalization of Eq.(\ref{SokolovTernov}) is \cite{bk68}
\begin{equation}
      W
       \ =\
      \frac{1}{2  \tau_{0}}
      \left[1-
      \frac{2}{9}(\hat{\xi}\cdot\hat{s})^{2}+
      \frac{8}{5\sqrt{3}}\,\hat{\xi}\cdot\hat{b}
          \right]
\end{equation}
where $\hat s$ = direction of motion and
$\hat b = ({\hat s} \times {\dot{\hat s}})/|{\dot{\hat s}}|$.
$\hat b$ is the magnetic field direction if the electric field vanishes
and the motion is perpendicular to the magnetic field.

The corresponding instantaneous rate of build-up of
                  polarization along $\hat \xi$ is
\begin{equation}
      \tau^{-1}_{\rm bk}
       \ =\
      \tau^{-1}_{0}
      \left[1-
      \frac{2}{9}(\hat{\xi}\cdot\hat{s})^{2}
          \right]
\label{eq:tauBK}
\end{equation}

\vspace{5mm}
\paragraph{The T-BMT equation}

Neglecting radiative spin flip,
                   the motion of the rest-frame spin
expectation value $\vec\xi$ of a relativistic charged particle
traveling in electric and magnetic fields
is governed by the Thomas-BMT equation
 $d{\vec\xi}/dt = \vec{\Omega} \times {\vec\xi}$
(Sec.2.7.1 in \cite{handbooka}).

%We write
%\begin{equation}
%      \vec{\Omega}
%       \ =\ 
%      \vec{\Omega}^{\rm ref}
%        +
%      \vec{\omega}^{\rm imp} + \vec{\omega}^{\rm sb}
%\end{equation}
%where $\vec{\Omega}^{\rm ref}$ is due to design fields on the design orbit 
%and
%$\vec{\Omega}^{\rm ref} + \vec{\omega}^{\rm imp} 
%(\equiv \vec{\Omega}^{\rm co})$ is
%due to fields on the closed orbit whereby
%              $\vec{\omega}^{\rm imp}$ is due to field
%imperfections and corrections.
%$\vec{\omega}^{\rm sb}$ is due to
%synchrotron and/or betatron motion with respect to the closed orbit.

We write
\begin{equation}
      \vec{\Omega}
       =
      \vec{\Omega}^{\rm co}
        +
      \vec{\omega}^{\rm sb}
\end{equation}
where $\vec{\Omega}^{\rm co}$ is due to the fields on the closed orbit, whence
$\vec \Omega^{\rm co}(s + C) = \vec \Omega^{\rm co}(s)$.
$\vec \Omega^{\rm co} = \vec \Omega^{\rm ref} + \vec \omega^{\rm imp}$, 
where $\vec \Omega^{\rm ref}$ contains the
design fields and $\vec \omega^{\rm imp}$ represents the effects of magnet
misalignments, correction fields etc. 
$\vec{\omega}^{\rm sb}$ is due to
synchrotron and/or betatron motion with respect to the closed orbit.

On the closed orbit the T-BMT equation
\begin{equation}
     \frac{d}{dt}
              \vec{\xi}
                          \ =\
          \vec{\Omega}^{\rm co}  \times
                                \vec{\xi}
\label{coSpin}
\end{equation}
can be solved in the form
\begin{equation}
         \vec{\xi}(s)
              \ =\
         {\bf{R}}_{3\times 3}^{\rm co}(s, s_{0})
         \vec{\xi}(s_{0})
\end{equation}
where ${\bf{R}}_{3\times 3}^{\rm co}$
is a rotation matrix.
The real unit eigenvector (rot. axis) for the one turn matrix
        ${\bf{R}}_{3\times 3}^{\rm co}(s+C, s)$,
denoted by
        $\hat{n}_{0}(s)$,
is the periodic spin solution on the closed orbit.
For a perfectly aligned flat ring with no solenoids,
        $\hat{n}_{0} (s) = \pm \hat{y}$.
The one turn matrix has a complex conjugate pair of
eigenvalues
$ e^{\pm i {2}\pi \nu_{\rm spin}}$.
Given $\hat{n}_{0}$,
                    we introduce a pair of unit vectors
$(\hat{m}_{0},\hat{l}_{0})$ 
such that
  $\hat{m}_{0} = \hat{l}_{0}\times\hat{n}_{0}$ and
  $\hat{l}_{0}=  \hat{n}_{0}\times\hat{m}_{0}$
fulfill Eq.(\ref{coSpin}), and such that
\begin{eqnarray}
       && \hat{m}_{0}(s_{0}+C)+
       i \hat{l}_{0}(s_{0}+C)
 \ =\
                        e^{i{2}\pi \nu_{\rm spin}}
             \left[
        \hat{m}_{0}(s_{0})+
       i\hat{l}_{0}(s_{0})
                            \right]
\end{eqnarray}
The $(\hat{m}_{0},\hat{l}_{0})$ 
are usually not periodic in $s$.
But by applying a further rotation
by an angle $\psi_{\rm spin}(s)$ around
        $\hat{n}_{0}$
we can construct the vectors
        $(\hat{m},\,
          \hat{l})$,
\begin{eqnarray}
       \hat{m}(s)+
       i\, \hat{l}(s)
 &=& e^{  -i \psi_{\rm spin}(s)}
             \left[
        \hat{m}_{0}(s)+
       i\hat{l}_{0}(s)
                            \right]
\end{eqnarray} 
By choosing
     $\psi_{\rm spin}(s+C)-\psi_{\rm spin}(s)=2\pi \nu_{\rm spin}$,
the set $(\hat{n}_{0}, \hat{m}, \hat{l})$
is then periodic in $s$ with period $C$. The vectors $(\hat{m}, \hat{l})$
are needed in Sec 2.7.8.

The closed orbit 
spin tune $\nu_{\rm spin}$ is the number of spin precessions per turn
around $\hat{n}_{0}$.
For a perfectly aligned flat ring without solenoids 
$\nu_{\rm spin} = a\gamma_0$,
where $a=(g-2)/2$ (see Sec.2.7.1 in \cite{handbooka})
and $\gamma_0$ is the Lorentz factor for the beam energy.
In this section and in Sec. 2.7.8 we use 
the symbol ``$a$'' instead of the symbol ``$G$'' used in the rest of the 
Handbook. Only the fractional part of the spin tune can be extracted from
the numerical values of the eigenvalues $ e^{\pm i {2}\pi \nu_{\rm spin}}$.
%For the definition of spin tune away from the closed orbit see \cite{mont98}.

\vspace{5mm}
\paragraph{The Baier-Katkov-Strakhovenko (BKS) equation}
Neglecting the effect of stochastic (synchrotron radiation) photon emission
on the orbit and imagining that all particles remain on the closed
orbit, the equation of motion for electron  polarization is \cite{bks70,stck69}
\begin{eqnarray}
      \frac{d\vec{P}}{dt}
       &=&
      {\vec{\Omega}}^{\rm co}\times\vec{P}
%Mods for killing splitting
%\\KILL
%        &-&
- 
%Mods for killing splitting
     \frac{1}{\tau_{0}(s)}
      \left[
         \vec{P}-\frac{2}{9}
         \hat{s}
         (
           \vec{P}\cdot\hat{s}
                           )
               +
      \frac{8}{5\sqrt{3}}
      {\hat{b}}(s)
          \right] 
%KILL  \nonumber
\end{eqnarray}
In the case of horizontal motion in a vertical magnetic field,
we have $\vec{\Omega} =  (a\gamma c/\rho)\hat{y}$,
and $\hat{b}(s) = \hat{y}$.

By integrating the BKS equation, one finds the
generalized Sokolov-Ternov formula for the asymptotic electron polarization
in arbitrary magnetic fields along  the closed orbit,
\begin{equation}
         \vec{P}_{\rm bks}
       \ =\
       -  \frac{8}{5\,\sqrt{3}}~ \hat{n}_0 ~
     \frac{
   {\oint ds
                \frac{
                        {\hat{n}_{0}}(s)
                              \cdot
                        {\hat{b}}(s)
                                }
                     {|\rho(s)|^{3}}
                                    }
                                      }
          {
   {\oint ds
                \frac{
                            \left[
                                 1-
                                 \frac{2}{9}
                       ({\hat{n}_{0}}(s)\cdot
                        \hat{s})^{2}
                                       \right]
                                                 }
                    {|\rho(s)|^{3}}
                                    }
                                     }
\label{GeneralizedST}
\end{equation}
See \cite{mont84} for a compilation of time scales.
Usually, in rings containing dipole spin rotators 
(Secs.2.7.3, 2.7.4 in \cite{handbooka})  the
polarization $|\vec{P}_{\rm bks}|$ cannot reach 0.9238 \cite{bar95a}.

The BKS polarization build-up rate is
\begin{equation}
      \tau^{-1}_{\rm bks}
       \ =\
      \frac{5\sqrt{3}}{8}
      \frac{r_{\rm e} \gamma^{5}\hbar}
           {m_{\rm e}}
      \frac{1}{C}\,
      \oint ds\,
                \frac{
                             \left[
                                 1-
                                 \frac{2}{9}\,
                       ({\hat{n}_0}\cdot
                        \hat{s})^{2}
                                       \right]
                                                 }
                     {|\rho(s)|^{3}}
\label{eq:tauBKS}
\end{equation}
This is in accord with Eq.(\ref{eq:tauBK})
by replacing $\hat \xi \rightarrow {\hat n}_0$ and averaging.

\vspace{5mm}
\paragraph{Radiative depolarization} The
stochastic element of photon emission
together with damping determines the equilibrium phase space
density distribution. The same photon emission also imparts a
stochastic element to $\vec{\omega}^{\rm sb}$  and then, via the T-BMT
equation, spin diffusion (and thus depolarization) can occur 
\cite{baiorl66}.
The  polarization is the result of a balance between the
Sokolov-Ternov effect and this radiative
depolarization. In the approximation that
the orbital motion is linear, the {\it value} of the polarization is
 essentially
the same at each point in phase space and azimuth and the polarization is 
aligned along
the Derbenev-Kondratenko vector $\hat n$ \cite{dk73}.

The unit vector field $\hat{n}$, which is also called the
{\it``invariant spin field''} \cite{mont98,beh2004,hvb99a,spin2000}, depends on $s$ and
$\vec u \equiv(x, p_x, y, p_y, z, \delta)$.
${\hat n}({\vec u};s)$ satisfies the T-BMT equation at
$({\vec u};s)$ and is periodic: $\hat{n}(\vec{u}; s)
                 =
            \hat{n}(\vec{u}; s+C)$.
On the closed orbit $\hat{n}(\vec{u}; s)$
                 reduces to $\hat{n}_{0}(s)$.

\vspace{5mm}
\paragraph{The Derbenev--Kondratenko--Mane formula}
Taking into account radiative depolarization
due to photon-induced longitudinal recoils, the equilibrium electron 
polarization along the $\hat{n}$ field is \cite{dk73,mane87a,mont98}
%KILL\newline
%KILL     ${P}_{\rm dk} &=&
%KILL        -\frac{8}{5\sqrt{3}}\times$
\begin{eqnarray}
%Insert inside
           {P}_{\rm dk} &=&
             -\frac{8}{5\sqrt{3}}
%Insert inside
     \frac{
   { \oint {ds} \left<  \frac{1}{|\rho(s)|^{3}}
              \hat{b}
              \cdot
      (
       \hat{n}-
                 \frac{\partial{\hat{n}}}
                      {\partial{\delta}}
                           )
          \right>_{s}
                                    }
                                      }
          {
   {\oint {ds} \left< \frac{1}{|\rho(s)|^{3}}
          ( 1-
              \frac{2}{9}
     { (
       \hat{n}\cdot\hat{s}
                           )}^{2}
              +
      \frac{11}{18}
      \left(
                 \frac{\partial{\hat{n}}}
                      {\partial{\delta}}
                                        \right)^{2} \, )
          \right>_{s}
                                    }
                                     }
\label{eq:PDK}
\end{eqnarray}
where         $<\ >_{s}$
denotes an average over phase space
at azimuth $s$.  This formula differs from
Eq.(\ref{GeneralizedST}) by the
inclusion of the terms with
 $\frac{\partial{\hat{n}}}{\partial{\delta}}$
and use of $\hat{n}$ instead of $\hat{n}_{0}$.
The ensemble average of the polarization is
\begin{equation}
  { \vec  P}_{\rm ens,dk}(s)
              \ =\
     P_{\rm dk}~
     \langle \hat{n} \rangle_{s}
\end{equation}
and $ \langle \hat{n}\rangle_{s}$ is very nearly aligned along
 ${{\hat n}_0}(s)$ (see the angle estimate below). The {\it value}
of the ensemble average,
 ${P}_{\rm ens,dk}(s)$, is essentially independent of $s$.

The effect of transverse recoil can also be included but contributes
derivative terms analogous to
 $\frac{\partial{\hat{n}}}{\partial{\delta}}$ which are typically
a factor $\gamma$ smaller than
 $\frac{\partial{\hat{n}}}{\partial{\delta}}$ and can  be
 neglected unless  $\frac{\partial{\hat{n}}}{\partial{\delta}}$
is very small \cite{bm88,hs87}. If 
 $\frac{\partial{\hat{n}}}{\partial{\delta}}$ were to vanish, a  $P_{\rm dk}$
of 99.2 \% could be reached \cite{bm88,hs87,mont98}.

In the presence of radiative depolarization Eq.(\ref{eq:tauBKS}) becomes
\begin{eqnarray}
      \tau^{-1}_{\rm dk}
       &=&
      \frac{5\sqrt{3}}{8}
      \frac{r_{\rm e} \gamma^{5}\hbar}
           {m_{\rm e}}
      \frac{1}{C}
%Mods for killing splitting
%\\KILL
%    &\times&  \oint ds
               \oint ds                             \left<
%Mods for killing splitting
                \frac{
                                 1-
                                 \frac{2}{9}
                       (\hat{n}\cdot
                        \hat{s})^{2}
                             +
                        \frac{11}{18}
                 \left(
                 \frac{\partial{\hat{n}}}
                      {\partial{\delta}}\right)^{2}
                                                 }
                     {|\rho(s)|^{3}}
                                       \right>_s
%KILL \nonumber
\end{eqnarray}
This can be written in the form:
\begin{eqnarray}
      \frac{1}{{\tau}_{\rm dk}}
           &=&
      \frac{1}{\tau_{\rm st}}
            +
      \frac{1}{\tau_{\rm dep}}\ ,
\end{eqnarray}
where
              $\tau_{\rm st}^{-1}$
can be (very well) approximated by $\tau_{\rm bks}^{-1}$ in (15) and
\begin{eqnarray}
      \tau^{-1}_{_{\rm dep}} 
      &=&
      \frac{5\sqrt{3}}{8}
      \frac{r_{\rm e}\gamma^{5}\hbar}
           {m_{\rm e}}
      \frac{1}{C}
%\nonumber \\
%      &\times&  \oint ds\,
                 \oint ds\,
                            \left<\,
                \frac{
                        \frac{11}{18}\,
                 \left(
                \frac{\partial{\hat{n}}}
                      {\partial{\delta}}\right)^{2}}
                     {|\rho(s)|^{3}}
                                       \right>_s 
\end{eqnarray}
The time dependence for build-up
from an initial polarization $P_0$ to
equilibrium is
\begin{equation}
P(t)\ =\
  {P}_{\rm ens,dk}
    \left[
          1- e^{-t/{\tau_{\rm dk}}}
              \right] +
                        P_0 e^{-t/{\tau_{\rm dk}}}
\end{equation}
This formula can be used to calibrate polarimeters 
(see Eqs.(21) and (22), Sec.2.7.8)
\cite{hpol2a}.
However, the calibration will be imprecise if 
$\frac{\partial{\hat{n}}}{\partial{\delta}}$
in the numerator of Eq.(\ref{eq:PDK}) is not well enough known.
For examples of build-up curves see \cite{bar95a}.

\vspace{5mm}
\paragraph{Resonances}
Away from the spin--orbit
resonances \footnote{In fact the resonance condition should be more 
precisely expressed in terms of the so--called amplitude dependent spin 
tune \cite{mont98,beh2004}. But for typical electron/positron rings the amplitude 
dependent spin tune   differs only insignificantly from $\nu_{\rm spin}$.}
(see also Eq.(11), Sec.2.7.8)
\begin{equation}
       \nu_{\rm spin}
        \ =\
          k_{0}
          +
          k_{x}\nu_{x}
          +
          k_{y}\nu_{y}
          +
          k_{z}\nu_{z}
\end{equation}
$\hat{n}(\vec{u}; s) \approx {\hat n}_0(s)$. But near
resonances $\hat{n}(\vec{u}; s)$ deviates from ${\hat n}_0(s)$
by typically tens of milliradians at a few tens of GeV and the
deviation increases with distance in phase space  from the closed orbit.
The {\it ``spin--orbit coupling function''}
                $\frac{\partial{\hat{n}}}
                      {\partial{\delta}}$, which quantifies the depolarization,
can then be large and the equilibrium polarization can then be small.
Note that even very close to resonances,
   $|\langle \hat{n}\rangle_{s}| \approx 1$:
the ensemble average
polarization is mainly influenced by the value of
    $P_{\rm dk}$ in Eq.(\ref{eq:PDK}).

To get high polarization, one must have $(\partial{\hat{n}}/
                      \partial{\delta} )^{2} \ll 1$
in dipole magnets.  The machine optimization required to make
$\frac{\partial{\hat{n}}} {\partial{\delta}}$
                    small  is called {\it ``spin matching''}
(Sec.2.7.8).

\vspace{5mm}
\paragraph{Asymmetric wigglers}
If  $\tau^{-1}_{\rm bks}$ is very low because the energy is low
and/or the average curvature is small the polarization rate can be enhanced 
(see Eq.(\ref{eq:tauBKS}))
by installing an {\it ``asymmetric wiggler''}, i.e. 
a string of dipoles in which short dipoles with  high fields 
are interleaved with long dipoles with low fields of opposite polarity
while ensuring that
the field integral of the string vanishes. For more details, and discussion
of advantages and disadvantages see \cite{mont84}. A particular
potential disadvantage is that the enhanced radiation loss can require 
that  extra rf power be installed and that the energy spread increases
so that the depolarization rate increases owing  to stronger  synchrotron
 sideband resonances (Sec.2.7.8).

\vspace{5mm}
\paragraph{Kinetic polarization}
The (numerator) term linear in  $\frac{\partial{\hat{n}}}{\partial{\delta}}$
in Eq.(\ref{eq:PDK})
 is due to a correlation between the spin orientation and the 
radiation power \cite{mont84}. In rings where ${\hat n}_0$ is horizontal due,
 say, to the
presence of a solenoid Siberian Snake (Secs.2.7.3, 2.7.4 in \cite{handbooka})
 \cite{spin96a}, 
$\frac{\partial{\hat{n}}}{\partial{\delta}}$ has a vertical component
in the dipole fields. This can lead to a build-up of polarization
({\it ``kinetic polarization''}) even though
the pure Sokolov--Ternov effect vanishes. The rate is $\tau^{-1}_{\rm dk}$. 

\vspace{5mm}
\paragraph{Phase space and polarization evolution equations}
If the orbital phase space density $\psi$ obeys an equation of the 
Fokker--Planck type (Sec.2.5.4 in \cite{handbooka})
\begin{eqnarray}
\frac{\partial \psi} {\partial s}   &=&
       {\cal L}_{{}_{\rm FP}} \; \psi 
\end{eqnarray}
where ${\cal L}_{{}_{\rm FP}}$ is the orbital Fokker--Planck operator,
then the spin diffusion is described by the ``Bloch'' equation
\begin{eqnarray}
\frac{\partial\vec{\cal P}} {\partial s}   &=&
     {\cal  L}_{{}_{\rm FP}} \; \vec{\cal P} +
\vec{{\bar \Omega}} \times \vec{\cal P} 
\end{eqnarray}
where $\vec{{\bar \Omega}} = {\vec \Omega}/{(ds/dt)}$ 
and $\vec{\cal P}$ is the 
{\it ``polarization density''} $\equiv 2/\hbar \times$(density in phase space
per particle of spin angular momentum)~\cite{kh97,dbkh98}.
To include the Sokolov--Ternov effect see \cite{dk75}. 

\vspace{5mm}
\paragraph{Beam energy calibration}
A polarized electron beam can be depolarized by applying a weak oscillating 
magnetic field perpendicular to $\hat n_0$ with a frequency $f_{\rm rf}$
related to the fractional part of the spin tune $\tilde \nu_{\rm spin}$ by
\begin{eqnarray}
f_{\rm rf}&=&f_{\rm c}\tilde \nu_{\rm spin} \qquad \!\!\!\! {\rm or} \; \; \;
f_{\rm rf}\ =\ f_{\rm c}(1 - \tilde \nu_{\rm spin})
\end{eqnarray}
where $f_{\rm c}$ is the circulation frequency of the beam \cite{symor}. 
Thus the required  $f_{\rm rf}$ gives an accurate
measurement of $\tilde \nu_{\rm spin}$ and this gives high relative precision
knowledge of $\nu_{\rm spin}$. By relating $\nu_{\rm spin}$  to the average
energy of each beam, high precision measurements of the centre--of--mass 
energy of colliding $e^+$--$e^-$ beams  and of the masses of vector mesons 
such as the $\Upsilon$ family and the $Z$ can be obtained \cite{shat89,lep95,
plac98,bar84,
han84}. Other beam parameters can also be measured \cite{bar85a}.
The polarization need not be large for these measurements so that by Eq.(21)
the depolarization can be repeated at intervals of about $\tau_{\rm dk}$.

\vspace{3mm}
\paragraph{Concluding remarks}
For an overview of measurements see \cite{bar96,bar95a,shat90}. For an overview
of
the theoretical background see \cite{mont98}.

\newpage

\setcounter{equation}{0}

\section*{\large \bf {2.7.8.~Computer Algorithms and Spin Matching}}

There are two classes of computer algorithm for estimating the
equilibrium polarization in real rings:

\begin{itemize}
\item[(i)] Methods based on evaluating
   $\frac{\partial{\hat{n}}} {\partial{\delta}}$ in the 
Derbenev--Kondratenko--Mane (DKM) formula (Eq.(16) of Sec.2.7.7)
given the ring layout and magnet strengths; and

\item[(ii)] The SITROS \cite{kew89} and SLICKTRACK \cite{bar2005} algorithms which estimate $\tau_{\rm dep}$ 
(Sec.2.7.7)
               using  Monte--Carlo tracking.
\end{itemize}

The class (i) algorithms are further divided according
to the degree of linearization of the spin and orbital motion:
\begin{itemize}
\item
[(ia)] The SLIM family (SLIM \cite{chao81,chao2}, SLICK \cite{bar82},
 SITF \cite{kew89})  and
     SOM \cite{yok96} and ASPIRRIN \cite{spin96b}. The latter two utilize the  
   {\it ``betatron--dispersion''} formalism outlined below and all
     are based on a linearization of the orbital and spin motion.
\item
[(ib)] SMILE \cite{mane87b}: Linearized orbital motion but nonlinear spin motion;
\item
[(ic)] SODOM \cite{yok92}: Linearized orbital motion but nonlinear spin motion;
\item 
[(id)] SpinLie: Nonlinear orbital motion and nonlinear spin motion
(Sec.2.7.9 in \cite{handbookb}); and
\item
[(ie)] SPRINT \cite{hh96,gh2000}: Linearized orbital motion but nonlinear spin motion.
\end{itemize}
\vspace{5mm}
\paragraph{The linear approximation -- SLIM}
We now present expressions for
$\frac{\partial{\hat{n}}} {\partial{\delta}}$
in an approximation in which the orbit
and spin motion are linearized
and in which
$\vec{\omega}^{\rm sb}$ (Sec.2.7.7)
is linearized as in Eq.(2) below (the SLIM formalism).
In linear approximation we write (see Sec.2.7.7)
\begin{equation}
            \hat{n}(\vec{u}; s)
                 \ =\
            \hat{n}_{0}(s)
                  +
            \alpha(\vec{u}; s)\hat{m}(s)
                  +
            \beta(\vec{u}; s)\hat{l}(s)
\end{equation}
valid for
$\sqrt{\alpha^{2}+\beta^{2}}\ll 1$ and we write the components
          $\omega_{z}^{{\rm sb}}$,\,
          $\omega_{x}^{{\rm sb}}$,\,
          $\omega_{y}^{{\rm sb}}$
in the form \cite{mr83,bhr1}
\begin{equation}
       \left( \begin{array}{c}
                \omega_{z}^{{\rm sb}}      \\ \omega_{x}^{{\rm sb}}
                                      \\ \omega_{y}^{{\rm sb}}
                \end{array}
         \right)
               \ =\
        \bf{F}_{3\times 6}
       \left( \begin{array}{c}
               {x}       \\{p}_{x}
          \\   {y}       \\ {p}_{y}
          \\   {z}       \\ \delta
                \end{array}
         \right)
\end{equation}
where $\vec{u} \equiv  ({x},{p}_{x},
                {y},{p}_{y},{z},\delta)$
describes motion with respect to the closed orbit.
In particular ${p}_{x} = {{x}}'$ and 
${p}_{y} = {{y}}'$\,(except in solenoids).

The detailed forms of the matrix $\bf{F}_{3\times 6}$ for
         bending magnets,
         quadrupoles,
         skew quadrupoles,
         solenoids
     and rf cavities
       can be found in \cite{bhr1}.
The orbit motion in sextupoles is linearized.
For example for a quadrupole,
defining
$\tilde{g}=-(1+a\gamma_0)\,g$
where $g=\frac{e}{p_0}\frac{\partial B_y}{\partial x}$
one has
\begin{eqnarray}
      {\bf{F}}(s)
                             &=&                            
      \left( \begin{array}{cccccc}
          0 & 0 & 0 & 0  & 0 & 0 \\
0 & 0 & \tilde{g} & 0 & 0 & 0   \\  
\tilde{g} & 0 & 0 & 0 & 0 & 0          
              \end{array}
       \right)
\end{eqnarray}

In linear  approximation the combined orbit and spin motion is
described by 8 $\times$
8 transport matrices of the form
\begin{equation}
   \bf{\hat{M}} \ =\ \left( \begin{array}{rr}
                \bf{M}_{6\times 6}
             &  \bf{0}_{6\times 2}  \\
                \bf{G}_{2\times 6}
             &  \bf{D}_{2\times 2}
              \end{array}
       \right)
\end{equation}
acting on the vector  $(\vec{u},  \,  \alpha,\,\beta)$,
where  $\bf{M}_{6\times 6}$
is a symplectic matrix describing orbital motion and 
               $\bf{G}_{2\times 6}$ 
describes the coupling of the spin variables $(\alpha,\, \beta)$
to the orbit and depends on  $\hat{m}(s)$ and $\hat{l}(s)$ (see e.g. Eq.(14)).
$\bf{D}_{2\times 2}$
is a rotation  matrix associated with the spin basis rotation of 
Eq.(12) in Sec.2.7.7 \cite{mr83,bhr1}.

The eigenvectors for one turn defined by
$        {\bf{\hat{M}}}(s_{0}+C,s_{0})\cdot\vec{q}_{\mu}=
         \hat{\lambda}_{\mu}\cdot\vec{q}_{\mu}$
are  written in the form
\begin{eqnarray}
                \vec{q}_{k}(s_{0})
               &=&
       \left( \begin{array}{c}
                \vec{v}_{k}(s_{0})
                       \\ \vec{w}_{k}(s_{0})
                \end{array}
         \right) , \quad
                \vec{q}_{-k}(s_{0})
                =
         [
                \vec{q}_{k}(s_{0})
                                  ]^{*}
\nonumber \\
\nonumber \\
  && \hspace{1cm}{\rm for} \ k = I,\ II,\ III \ ;
\nonumber \\
\nonumber \\
                \vec{q}_{k}(s_{0})
               &=&
       \left( \begin{array}{c}
                \vec{0}_{6}(s_{0})
                       \\ \vec{w}_{k}(s_{0})
                \end{array}
         \right) \ ,\ \
                \vec{q}_{-k}(s_{0})
                =
         [
                \vec{q}_{k}(s_{0})
                                  ]^{*}
\nonumber\\
\nonumber\\
 && \hspace{1cm}{\rm for}\ k=IV
\end{eqnarray}
for arbitrary $s_0$.
The  $\vec{v}_{k}$
are the eigenvectors for orbital motion with eigenvalues
$ {\lambda}_k = e^{\textstyle -i {2}\pi{\nu}_{k}}$
and with $\nu_{-k} = -\nu_{k}$ $(k=I,\ II,\ III)$. These eigenvectors
obey the orthogonality relations, and have the normalization of \cite{chao81}.
The corresponding eigenvalues of
${\bf{\hat{M}}}(s_{0}+C,s_{0})$ are
$\hat{\lambda}_{k}=\lambda_{k}$ $(k=I,\ II,\ III)$
and
$\hat{\lambda}_{_{IV}}= e^{\textstyle -i {2}\pi{\nu}_{_{IV}}}$
with $\nu_{_{IV}}=\nu_{\rm spin}$ and with $\nu_{_{-IV}} = -\nu_{_{IV}}$.

The spin parts of the eigenvectors
       $\vec{w}_{k}(s_{0})$ \ $(k=I,\ II,\ III)$ and
       $\vec{w}_{_{IV}}(s_{0})$ can be written as
\begin{eqnarray}
   \vec{w}_{k}(s_{0}) &=& -\left[
            {\bf{D}}(s_{0}+C,s_{0})
           -\hat{\lambda}_{k}
                       \right]^{-1}
%Mods for killing splitting
%\nonumber \\ KILL
%      &\times&     {\bf{G}}(s_{0}+C,s_{0})
                    {\bf{G}}(s_{0}+C,s_{0})
%Mods for killing splitting
            \vec{v}_{k}(s_{0})
 \nonumber\\
 \nonumber\\
 && \hspace{1cm} \mbox{for}\
           k=I,\ II,\ III\ ;
\nonumber\\
\\
                \vec{w}_{_{IV}}(s_{0})
               &=&
           \frac{1}{\sqrt{2}}
       \left( \begin{array}{c}
                1
            \\ -i
                \end{array}
         \right)
                        e^{
          -i\, \psi_{\rm spin}(s_{0})}
\nonumber \\
\nonumber \\
  && \hspace{1cm} \mbox{for}\
     k= IV
\nonumber
\end{eqnarray}
and
\begin{eqnarray}
    \vec{w}_{-k}(s_{0})\ =\
          \left[
    \vec{w}_{k}(s_{0})
                          \right]^{*} ,
   (k=I, II,  III, IV)
\nonumber
\end{eqnarray}

In this linear approximation
           $\hat{n}(\vec{u}; s)$
can be obtained via \cite{mane85125,bhrnotes,bhr1}
\begin{eqnarray}
      &&    \hat{n}(\vec{u}; s)
                  -
           \hat{n}_{0}(s)
             \ \equiv
      \ \left( \begin{array}{c}
           \alpha(\vec{u}; s)
    \\     \beta(\vec{u}; s)
                \end{array}
         \right)
%Mods for killing splitting
%\nonumber\\ KILL
%         &=&
=
                               \sum_{k=I,II,III}
%Mods for killing splitting
        \left\{
               A_{k}
                         \vec{w}_{k}(s)
              +A_{-k}
                         \vec{w}_{-k} (s)
        \right\}
\label{eq:7}
\end{eqnarray}
where the amplitudes
         $A_{k}$
are determined by the orbit via
\begin{equation}
               {\vec{u}}\, (s)
                        \ =\
                               \sum_{k=I,II,III}
        \left\{
               A_{k}  \vec{v}_{k} (s)
              +A_{-k} \vec{v}_{-k} (s)
        \right\}
\end{equation}
Then with respect to the $(\hat{n}_{0}, \hat{m}, \hat{l})$ frame,
\begin{eqnarray}
                 \frac{\partial{\hat{n}}}
                      {\partial{\delta}}
                        &\equiv&  i \sum_{k=I,II,III}
        \left\{
               v_{k5}^{*}
                         \vec{w}_{k}
                       -
               v_{k5}
                         \vec{w}_{k}^{*}
        \right\}
\nonumber\\
\nonumber\\
                        &=&
              -2~
                          {\rm Im}
                \sum_{k=I,II,III}
               v_{k5}^{*}    \vec{w}_{k}
\end{eqnarray}
Note that this is independent of the phase space vector
$\vec{u}$ and that 
$\frac{\partial{\hat{n}}} {\partial{\delta}}$ is periodic in azimuth in the
machine coordinate system.
In this approximation the depolarization time is then (Eq.(20), Sec.2.7.7)
\begin{eqnarray}
      \tau_{\rm dep,lin}^{-1}
               &=&
      \frac{55\sqrt{3}}{36}
      \frac{r_{\rm e} \gamma_0^{5}\hbar}
           {m_{\rm e}}
      \frac{1}{C} \int_{s_{0}}^{s_{0}+C}{d\tilde{s}}
                 \frac{1}{|\rho(\tilde{s})|^{3}}
%Mods for killing splitting
%\nonumber \\ KILL
%      &\times&   \sum_{\mu=1}^{2}
                  \sum_{\mu=1}^{2}
%Mods for killing splitting
         \left({\rm Im}\sum_{k=I - III}
              [v_{k5}^{*}
                            (\tilde{s})
                          w_{k\mu}
                            (\tilde{s})
                                       ]
         \right)^{2}
\nonumber \\
\end{eqnarray}

This is the formula used in SLIM to calculate the depolarization rate.
SLIM is based on thin lens optics. SLIM--like programs for thick lens
optics are SLICK  and SITF. Each term in Eq. (9) is basically the product of
the sensitivity of an orbit amplitude to a change of $\delta$ and
the sensitivity of $\hat{n}$ to a change of that orbit amplitude.

Using the  6 $\times$ 6 symplectic unit matrix $\bf S$  defined in 
\cite{chao81} and the relation 
$A_{k} = - i \vec{v}^{\dagger}_{k} {\bf S} \vec{u}$,
%To display the explicit dependence of $\hat n$ on $\vec u$,
~Eq. (\ref{eq:7}) can be written  to display the explicit dependence of 
$\hat n$ on $\vec u$  as
\begin{eqnarray}
      &&  
      \ \left( \begin{array}{c}
           \alpha(\vec{u}; s)
    \\     \beta(\vec{u}; s)
                \end{array}
         \right)
%Mods for killing splitting
%\nonumber\\ KILL
%         &=&
=
                 2 ~{\rm Im} \left\{ \sum_{k=I,II,III}
%Mods for killing splitting
              ~\vec{w}_{k}(s) \cdot \vec{v}^{\dagger}_{k}(s) {\bf S} \right\}
                         \vec{u}(s) = {\bf{H}_{2\times 6}} ~\vec{u}(s)
\nonumber
\end{eqnarray}

In this linearized theory the vectors $\hat{n}(\vec{u}; s)$ and
$\frac{\partial{\hat{n}}} {\partial{\delta}}$ display only first
order resonance behaviour, namely the resonances
\begin{equation}
       \nu_{\rm spin}
         \ =\
          k_{0}
          +
          k_{I}\nu_{I}
          +
          k_{II}\nu_{II}
          +
          k_{III}\nu_{III}
\end{equation}
with
$|k_{I}| + |k_{II}| + |k_{III}| =1$. 
They arise from the denominator
matrix in Eq.(6). The  theory is not valid beyond the limit
$\sqrt{\alpha^{2}+\beta^{2}}\ll 1$. 

In this formalism the horizontal and longitudinal orbital variables 
are usually coupled. See, for example, the symbolic forms of the orbital 
eigenvectors under ``{\it Harmonic closed orbit spin matching}'' below.
However, the eigentunes are usually very close to those associated 
with pure transverse ($x, y$) and longitudinal ($s$) motion so that 
in the absence of $x-y$ coupling one can often
make the associations: $I \rightarrow x$, $II \rightarrow y$ and
$III \rightarrow s$.

This  formalism forms the natural language for the method of maximizing
the polarization called {\it ``spin matching''}.
Thus comments on the other programs will be postponed until later.

\vspace{5mm}
\paragraph{Spin matching in the SLIM formalism}
In practice the spin matching of real rings takes place in stages as follows.
\\
\\
\noindent
Stage 1: {\it Strong synchrobeta spin matching of the perfectly aligned ring} 
\\
\\
    From Sec.2.7.7 it is clear that to maximize the polarization we must 
minimize $\tau_{\rm dep}^{-1}$. Then by 
Eq.(10) 
we need to minimize
$v_{k5}^{*}$\ \ $(k=I,\ II,\ III )$ or the components of $\vec{w}_{k}$
at azimuths where $1/|\rho(s)|^{3}$ is large.
The $v_{k5}^{*}$ determine the orbit excitation due to synchrotron
radiation (Sec.2.1.4 in \cite{handbookb}) \cite{chao75}. 
 In particular, for rings without $x-y$ 
 coupling,
$v_{_{II5}}^{*}$ usually vanishes in the arcs since the
vertical dispersion $\eta_{y}$ vanishes. However,
$v_{_{II5}}^{*}$ does not vanish inside spin rotators 
(Secs.2.7.3, 2.7.4 in \cite{handbookb})
 containing
vertical bends. On the other hand
$v_{_{I5}}^{*}$ tends not to vanish in the arcs since the  horizontal
dispersion $\eta_{x} \neq 0$. Finally, $v_{_{III5}}^{*}$ essentially never 
vanishes.
Each case must be evaluated individually but the minimal recipe is to try to
minimize $\vec{w}_{k}$ for ($k=I,\ II,\ III $)
only at azimuths where $|v_{_{k5}}(s)|^{2}/|\rho(s)|^{3}$ is sufficiently
 large.
This in turn requires (Eq.(6)) that ${\bf{G}}(s+C,\,s)\cdot \vec{v}_{k}(s)$ for
($k=I,\ II,\ III $) be minimized.
This must be  achieved by designing the ring layout with this in mind
and then providing sufficient flexibility in the optics by
providing enough independently powered quadrupoles.
Subsequent calculations with SLIM will indicate whether the match criteria
for the adopted design suffice. 

Consider, for example, a specific mode, $k$.
Label  those bending magnets at which
$|v_{_{k5}}(s)|^{2}/|\rho(s)|^{3}$
is large by
  ${\mu_{1}}^{(k)},\,
   {\mu_{2}}^{(k)},\,
   \cdot
   \cdot
   \cdot,\,
   {\mu_{n_{k}}}^{(k)}$.
Then the suppression of depolarization associated with the $k$th mode
requires that
$\vec{w}_{k}(s_{\mu_{i}}) = 0$ for all ($i = 1$ to $n_{k}$).
In general (see Eq.(6)) this in turn requires \cite{mr84}
\begin{eqnarray}
      {\bf{G}}(s_{\mu_{2}},\,s_{\mu_{1}})\,
      \vec{v}_{k}(s_{\mu_{1}})
           &=&
      0
\nonumber
\\
      {\bf{G}}(s_{\mu_{3}},\,s_{\mu_{2}})\,
      \vec{v}_{k}(s_{\mu_{2}})
           &=&
      0
\nonumber
\\
\vdots
\nonumber
\\
      {\bf{G}}(s_{\mu_{1}}+C,\,s_{\mu_{n}})\,
      \vec{v}_{k}(s_{\mu_{n}})
           &=&
      0
\end{eqnarray}
where we suppressed the superscript label ``$k$''.
To fulfill Eq.(12) we then require the\,
      $G_{ij}(s_{\mu_{l+1}},\, s_{\mu_{l}})$
to vanish when the $j$th component of
      $\vec{v}_{k}$
does not vanish. 
The matrix
         $\bf{G}$ can be written in the form
\begin{eqnarray}
        {\bf{G}}(s_{2},\,s_{1})
                  &=&
           \int_{s_{1}}^{s_{2}}{d\tilde{s}}~
                {\bf{D}}(s_{2},\,\tilde{s})\,
                {\bf{G}}_{0}(\tilde{s})\,
                {\bf{M}}(\tilde{s},\,s_{1})
\nonumber
\end{eqnarray}
where
\begin{equation}
        {\bf{G}}_{0}
                  \ =\
      \left( \begin{array}{rrr}
          l_{s}  & l_{x}  &
           l_{y}               \\
         -m_{s}  & -m_{x} &
         -m_{y}
             \end{array}
      \right)\cdot
     {\bf{      {F}}}
\end{equation}
Thus  $G_{ij}(s_{\mu_{l+1}},\, s_{\mu_{l}})$
depends on the orientation of the
      $(\hat{m},\,\hat{l})$ vectors 
so that in some cases some elements of
      $G_{ij}(s_{\mu_{l+1}},\, s_{\mu_{l}})$
vanish automatically. But in general these conditions can only be
fulfilled by
adjusting quadrupole strengths --- while maintaining other necessary
features of the orbital optics.
 We call this {\bf strong  synchrobeta spin matching}.
A section of the ring satisfying a condition in Eq.(12) is 
{\it ``spin transparent''}~ for mode $k$.
The interpretation is immediate: the 
overall spin--orbit coupling for the section vanishes for mode $k$. 
Clearly, the exact spin matching conditions are very dependent
on the layout of a machine
and each case must be handled individually.
In thin lens approximation
the
      $\bf{G}$ matrix
for a quadrupole of length $l_{\rm q}$ is
\begin{eqnarray}
      \bf{G}
                      &=&
      \left( \begin{array}{llllll}
          -\tilde{q}l_{y}  & 0 & -\tilde{q}l_{x} & 0 & 0 & 0  \\
          +\tilde{q}m_{y}  & 0 & +\tilde{q}m_{x} & 0 & 0 & 0  
              \end{array}
       \right) 
\ \ \ \ \ \
\end{eqnarray}
where $\tilde{q}=(1+a\gamma_0)\,g\,l_{\rm q}$.
The thin and thick lens forms of $\bf G$ for other magnet types
are given in \cite{chao81,chao2,bar85b}.

If the  ${{G_{ij}}}(s_{\mu_{l+1}},\, s_{\mu_{l}})$ cannot be brought to zero
while maintaining an acceptable optic, then the 
${\bf{G}}(s_{\mu_{{l}}}+C,\,s_{\mu_{{l}}})\cdot \vec{v}_{k}(s_{\mu_{{l}}})$ 
themselves should be minimized.
This essentially means that the effects of elements
of the $\bf G$ matrices of sections of the ring are made to partially cancel 
one another. The spin matching of a ring with a solenoid Siberian Snake
(Secs.2.7.3, 2.7.4 in \cite{handbookb})
 has  provided an example of this \cite{spin96b}.
By Eq.(7) reduction of  
${\bf{G}}(s+C,\,s)\cdot \vec{v}_{k}(s)$ for ($k=I,\ II,\ III $) also  
reduces the angle between $\hat{n}$ and $\hat{n}_{0}$ at azimuth $s$.
\\
\\
\noindent
Alternative Stage 1: {\it Harmonic synchrobeta spin matching of the perfectly
aligned ring} 
\\
\\
If the strong spin matching methods just described are impractical
for some reason,
another approach aimed at minimizing the strengths of 
depolarizing resonances can be adopted.

Rewrite Eq.(6) as
\begin{eqnarray}
&&           
    \left[\,
      w_{k1}(s_{0}){\mp}i\,w_{k2}(s_{0})\,
                      \right]
%Mods for killing splitting
%                   \nonumber\\ KILL
%&=&
\ =\
%Mods for killing splitting
         -
   \frac{
         e^{{\pm}i
             \psi_{\rm spin}(s_{0}+C)}
                                   }
   {
   \left[
      e^{{\pm}i 2\pi\nu_{\rm spin}}
     -e^{-i 2\pi\nu_{k}}
                                 \right]
                                              }
%Mods for killing splitting
%\nonumber  \\ KILL
%& \times &
%Mods for killing splitting
           \int_{s_{0}}^{s_{0}+C}{d\tilde{s}}
      j_{k}^{({\mp})}(\tilde{s})
      e^{-i 2\pi\left[\nu_{k}{\pm}\nu_{\rm spin}\right]\tilde{s}/C
                                                 }
\nonumber
\end{eqnarray}
with
\begin{eqnarray}
     j_{k}^{({\mp})}(\tilde{s})
        &=&
      e^{{\pm}i
             [2\pi\nu_{\rm spin}\tilde{s}/C
                                    -\psi_{\rm spin}(\tilde{s})]}
%Mods for killing splitting
%\nonumber\\ KILL
%     && \times    \left( \begin{array}{ccc}
                  \left( \begin{array}{ccc}
%Mods for killing splitting
\!\!      l_{s}{\pm}im_{s}   &
      l_{x}{\pm}im_{x}   &
      l_{y}{\pm}im_{y}
\!\!              \end{array}
       \right)
%Mods for killing splitting
%\nonumber\\ KILL
%    & \times &
%Mods for killing splitting
      {\bf{F}}
                \vec{v}_{k}(\tilde{s})
      e^{+i
                2\pi\nu_{k}\tilde{s}/C
                                                 }
\nonumber \\
          &=&
      j_{k}^{({\mp})}(\tilde{s}+C)
%Mods for killing splitting
%\nonumber    \\ KILL
%          &=&
=
%Mods for killing splitting
      \sum_{p=-\infty}^{+\infty}\,
           c_{kp}^{({\mp})}
      e^{+i
                2\pi p\tilde{s}/C
                                                 }
\nonumber
\end{eqnarray}
$\Longrightarrow$
\newline
\begin{eqnarray}
           c_{kp}^{({\mp})}
       \!\!   &=& \!\!
     \frac{1}{C}    \int_{0}^{C}{d\tilde{s}}
      e^{i
      2\pi\left[\,\nu_{k}{\pm}\nu_{\rm spin}-p\,\right]\tilde{s}/C
                                                 }
           e^{{\mp} i\psi_{\rm spin}(\tilde{s})}
%Mods for killing splitting
%\nonumber    \\ KILL
%& \times & \left( \begin{array}{ccc}
            \left( \begin{array}{ccc}
%Mods for killing splitting
      l_{s}{\pm}im_{s}   &
      l_{x}{\pm}im_{x}   &
      l_{y}{\pm}im_{y}
              \end{array}
       \right)
%Mods for killing splitting
%\nonumber     \\ KILL
%& \times & {\bf{F}}
            {\bf{F}}
                \vec{v}_{k}(\tilde{s})
%Mods for killing splitting
\nonumber
\end{eqnarray}
so that
\begin{eqnarray}
 &&  
       \left[\,
      w_{k1}(s_{0}){\mp}i\,w_{k2}(s_{0})\,
                      \right]
                  \ =\
         e^{{\pm}i \psi_{\rm spin}(s_{0})}\,
%Mods for killing splitting
%\nonumber \\ KILL
%&\times&
%Mods for killing splitting
i\frac{C} {2\pi}\,
      \sum_{p=-\infty}^{+\infty}
           c_{kp}^{({\mp})}
   \frac{
      e^{-i
        2\pi\left[\,\nu_{k}{\pm}\nu_{\rm spin}-p\,\right]{s_{0}}/C
                                                 }
                                                    }
        {
            \left[\,\nu_{k}{\pm}\nu_{\rm spin}-p\,\right]
                                                    }
\nonumber
\end{eqnarray}

The condition that
$\vec{w}_{k}(s_{\mu_{i}}) = 0$ for all ($i = 1$ to $n_{k}$)
is  now be replaced by
\begin{eqnarray}
&& \!\!\!\!   
   \left[\,
      w_{k1}(s_{\mu_{i}}){\mp}i\,w_{k2}(s_{\mu_{i}})\,
                      \right]
                   \ =\
         e^{{\pm}i
             \psi_{\rm spin}(s_{\mu_{i}})}\,
%Mods for killing splitting
%\nonumber\\ KILL
% &\times&
%Mods for killing splitting
  i\frac{C} {2\pi}\,
      \sum_{p=-\infty}^{+\infty}
           c_{kp}^{({\mp})}
   \frac{
      e^{-i
    2\pi\left[\,\nu_{k}{\pm}\nu_{\rm spin}-p\,\right]{s_{\mu_{i}}}/C
                                                 }
                                                    }
        {
            \left[\,\nu_{k}{\pm}\nu_{\rm spin}-p\,\right]
                                                    }
%Mods for killing splitting
%\nonumber     \\ KILL
%            &=&0
= 0
%Mods for killing splitting
\nonumber
\end{eqnarray}

Near to the resonance
      $\nu_{k}\pm
       \nu_{\rm spin}-\tilde{p}\,=\,0$\,
the sum over $p$ is dominated by the term
containing
      $c_{k \tilde{p}}^{({\mp})}$.
This corresponds to the spins' seeing a stationary
field  in the $(\hat{n}_{0},\, \hat{m}_{0},\, \hat{l}_{0})$ frame,
proportional to $c_{k \tilde{p}}^{({\mp})}$,
which rotates spins away from $\hat{n}_{0}$.
Note that $c_{k \tilde{p}}^{({\mp})}$ is independent of $s_{\mu_{i}}$.
Approximate spin matching can be achieved for all $s_{\mu_{i}}$
by adjusting the optics so that an appropriate set of the
  $c_{k \tilde{p}}^{({\mp})}$
are small. This is called
   {\bf harmonic synchrobeta spin matching}. See also \cite{mr84,rs85}.

On resonance $e^{i
        2\pi\left[\,\nu_{k}\pm\nu_{\rm sp}
                      -\tilde{p}\,\right]\tilde{s}/C
                                                 }
                   =
                    1$.
Then  the coefficients
       $c_{k\tilde{p}}^{(\mp)}$
take the form
\begin{eqnarray}
           c_{k\tilde{p}}^{(-)}
          &=&
     \frac{1}{C}
           \int_{0}^{C}{d\tilde{s}}
      e^{-i
            \psi_{\rm sp}(\tilde{s})}
\nonumber\\
 && \times   \left[ \begin{array}{ccc}
      l_{s}+im_{s}   &
      l_{x}+im_{x}   &
      l_{y}+im_{y}
              \end{array}
       \right]
 \nonumber \\
& & \times ~ {\bf{F}}
                \vec{v}_{k}(\tilde{s})
\hspace*{.5cm}
    \mbox{for}\ \
       \nu_{k}+\nu_{\rm sp}\,=\,\tilde{p}
\nonumber\\
           c_{k\tilde{p}}^{(+)}
          &=&
     \frac{1}{C}
           \int_{0}^{C}{d\tilde{s}}
      e^{+i
            \psi_{\rm sp}(\tilde{s})}
\nonumber\\
      && \times   \left[ \begin{array}{ccc}
      l_{s}-im_{s}   &
      l_{x}-im_{x}   &
      l_{y}-im_{y}
              \end{array}
       \right]
   \nonumber \\
&& \times ~ {\bf{F}}
                \vec{v}_{k}(\tilde{s})
\hspace*{.5cm}
    \mbox{for}\ \
       \nu_{k}-\nu_{\rm sp}\,=\,\tilde{p}
%\nonumber
\end{eqnarray}
 For mode $k$ and orbit amplitude $A_k$, 
the so-called {\em ``resonance strengths''} are   given by
 $A_{k} c^{-}_{k\tilde{p}}$ and  $A_{-k} (c^{+}_{k\tilde{p}})^*$.
The $c^{+}_{k\tilde{p}}$ and $c^{-}_{k\tilde{p}}$
can be obtained from the SLIM algorithm by calculating the matrix
$\bf{G}$ at the resonance for one turn but without the backward spin basis 
rotation (Eq.(12), Sec.2.7.7 in \cite{handbookb}) that, in SLIM, is applied at the end of one turn \cite{mr83}.
The concept of resonance strength (Eq.(2), Sec.2.7.5 in \cite{handbookb}) is important for the acceleration of
polarized protons. Normally only the case of flat rings with quadrupoles is considered so
that $\hat n_0$ is nominally vertical. The formalism presented here shows how to
define and easily obtain resonance strengths for each mode $k$ 
and in the presence of solenoids  and skew quadrupoles for arbitrary orientations of $\hat n_0$. 
See also \cite{lun97,bgb96,gh2000}.
\\
\\

\newpage

\noindent
{\it Reformulation in terms of beta functions and dispersion}\cite{mr86} 
\\
\\
We can reformulate Stage 1 by making a transformation
of the particle coordinates from 
$\vec u \equiv(x, p_x, y, p_y, z, \delta) $  to 
${\vec{\tilde u}} \equiv(\tilde x, {\tilde p}_x, \tilde y, {\tilde p}_y,
\tilde z, \delta)$
via the transformation
\begin{eqnarray}
    \vec{\tilde u} &=& {\bf K}\cdot \vec{u}
\nonumber
\end{eqnarray}
where
\begin{eqnarray}
      {\bf{K}(}s)
                             &=&                            
      \left( \begin{array}{cccccc}
          1 & 0 & 0 & 0  & 0 & -\eta_{1} \\
0 & 1 & 0 & 0 & 0 & -\eta_{2}   \\  
0 & 0 & 1 & 0 & 0 & -\eta_{3}  \\
0 & 0 & 0 & 1  & 0 & -\eta_{4} \\
\eta_{2} & -\eta_{1} & \eta_{4} & -\eta_{3} & 1 & 0  \\
0 & 0 & 0 & 0 & 0 & 1              
              \end{array}
       \right)
 \nonumber 
\end{eqnarray}
whereby the dispersion vector 
${\vec \eta} \equiv (\eta_1, \eta_2, \eta_3, \eta_4)$ is the periodic solution
of the linearized equations of motion for
$(x, p_x, y, p_y)$ with $\delta = 1$ and without the rf cavities. Then
with $\eta_{x} \equiv \eta_{1}, \eta_{y} \equiv \eta_{3}$
\begin{eqnarray} 
    \tilde x
            \ =\
           x-\delta \eta_{x},
\quad
   \tilde y
             \ =\
           y-\delta \eta_{y} \, .
\nonumber
\end{eqnarray}
The matrix $\bf K$ is symplectic so that the formalism remains canonical.
In particular, the new transfer matrices $\bf \tilde M$ and eigenvectors
$\vec{\tilde v}_{\mu}$ are obtained via
\begin{eqnarray} 
    {\bf \tilde M}(s_{2},s_{1}) 
  &=& {\bf  K}(s_{2})\cdot
 {\bf M}(s_{2},s_{1})\cdot
 {\bf K^{-1}}(s_{1})
\nonumber
\end{eqnarray}
and
\begin{eqnarray} 
    {\bf \tilde M}(s+C,s) 
  &=& {\bf  K}(s)\cdot
 {\bf M}(s+C,s)\cdot
 {\bf K^{-1}}(s)  \nonumber
\end{eqnarray}
\begin{eqnarray}
\Longrightarrow\ \
 {\vec{\tilde v}}_{\mu}(s)&=& {\bf K}(s)\,\vec{v}_{\mu}(s)
\nonumber
\end{eqnarray}
so that the eigenvalues and orthogonality conditions are unchanged. 
Furthermore the new matrices $\bf {\tilde F} $ and $\bf{\tilde G}$ are 

\begin{eqnarray} 
    {\bf \tilde F}(s) 
  &=&  {\bf F}(s)\cdot
 {\bf K^{-1}}(s)
\nonumber
\end{eqnarray}
and
\begin{eqnarray} 
    {\bf \tilde G}(s_2,s_1) 
  &=&  {\bf G}(s_2,s_1)\cdot
 {\bf K^{-1}}(s_1)  \nonumber
\end{eqnarray}

The depolarization rate then takes the form
\begin{eqnarray}
      \tau_{\rm dep,lin}^{-1}
              &=&
      \frac{55\sqrt{3}}{36}
%      \cdot
      \frac{r_{\rm e} \gamma_0^{5}\hbar}
           {m_{\rm e}}
      \frac{1}{C}\,
           \int_{s_{0}}^{s_{0}+C}{d\tilde{s}}
%\cdot
                 \frac{1}{|\rho(\tilde{s})|^{3}}
%Mods for killing splitting
%\nonumber\\ KILL
% &\times&
%Mods for killing splitting
                                    \sum_{\mu=1}^{2}
         \left(
                  {\rm Im}
                                    \sum_{k=I - III}
                      \left[\,f_{k}(\tilde s)
                {\tilde w}_{k\mu}
                            (\tilde{s})\,
                                \right]
         \right)^{2}
\nonumber
\end{eqnarray} 
with $f_{k}=
\sum_{n=1}^{6}\,
                [K^{-1}]_{5n}\cdot
{\tilde v}_{kn}^{*} = 
{v}_{k5}^{*}$             
and $\vec{\tilde{w}}_{k} = \vec{w}_{k}$. 
This formulation has the advantage that
in the special case, or the approximation,
of no orbital coupling, the 6 $\times$ 6 orbit matrices
just consist of three 2 $\times$ 2 matrices on the diagonal.
This is the case if there is no $x-y$ coupling and no dispersion
in the cavities.
Then we can make the identifications
\footnote{In the following we will choose the notations $(x,y,s)$ 
and $(I,II,III)$ according to the context. There should be no confusion.
If there is transverse--longitudinal coupling one can often still make
the associations $I \rightarrow x$, $II \rightarrow y$ and $III \rightarrow s$
just as when using the coordinates $u$.}:
$I \rightarrow x$, $II \rightarrow y$ and $III \rightarrow s$
and the eigenvectors $\vec{\tilde v}_{k}(s)$ of the revolution matrix can be
written in the form
\begin{eqnarray}
                \vec{\tilde v}_{I}
%               &=&
              \ =\
       \left( \begin{array}{l}
                \vec{t}_{x} \\ \vec{0}_{2} \\ \vec{0}_{2}
                \end{array}
         \right),
\
                \vec{\tilde v}_{II}
               \ =\
       \left( \begin{array}{l}
                                   \vec{0}_{2}\\
                \vec{t}_{y}\\
                                   \vec{0}_{2}\\
                \end{array}
         \right), 
\
%\nonumber\\
                \vec{\tilde v}_{III}
%             &=&
             \ =\
       \left( \begin{array}{l}
                                   \vec{0}_{2}   \\
                                   \vec{0}_{2}   \\
                \vec{t}_{z}
                \end{array}
         \right) \ ;
\nonumber\\ 
\nonumber\\
\nonumber\\
              \vec{t}_{r}
              \ =\
        \frac{1}{\sqrt{2\beta_{r}(s)}}
       \left( \begin{array}{c}
                \beta_{r}(s)
             \\ -[\alpha_{r}(s)+i]
                \end{array}
         \right)
                        e^{
          -i \psi_{r}(s)
                     }
\nonumber
\end{eqnarray}
$(r\equiv x,y,z)$ and the  $f_{k}$ are given by
$
   f_{I}\equiv f_{x}
      =
%    \cases{
 -(  \tilde{v}_{I1}\eta_{2}
           -
    \tilde{v}_{I2}\eta_{1})\ ;
  f_{II}\equiv f_{y}
      =
%    \cases{
 -(  
    \tilde{v}_{II3}\eta_{4}
          -
    \tilde{v}_{II4}\eta_{3})
$
and 
$
   f_{III}(s)\equiv f_{z}
      =
    \sqrt{
          \frac{\beta_{z}}{2}
                                   }
       e^{-i\psi_{z}(s)
                     }
$. 
The 
$|f_{{x}}|^2$ and $|f_{{y}}|^2$
are just the factors
\begin{eqnarray*}
     \frac{
           {\eta_{r}}^2 +\,{(\alpha_{r}\eta_{r}
                        +\beta_{r}\eta_{r}')}^2
                              }
          {2 {\beta_{r}}}\ \ (r\,=\,x,\,y)
\end{eqnarray*}
used in
\cite{sands70} to calculate emittances in the absence of transverse coupling.\,
In practice  $|f_{_{III}}|^2$ is  almost independent of $s$ since 
$\beta_{s}(s)$  is almost independent of $s$ (see below).
Note that these $\alpha$ and $\beta$ are Courant--Snyder parameters and
should not be confused with the quantities in Eq.(1).
With these coordinates the
       $\bf{\tilde F}$ matrix
for a quadrupole takes the form
\begin{eqnarray}
      \bf{\tilde{F}}
                      &=&
      \left( \begin{array}{cccccc}
          0 & 0  & 0 & 0 & 0 & 0 \\
          0 & 0  & \tilde{g} & 0 & 0 & \tilde{g}\eta_{3}   \\
          \tilde{g} & 0 & 0 & 0 & 0 & \tilde{g}\eta_{1}\
              \end{array}
       \right)
\nonumber
\end{eqnarray}
We can write
\begin{eqnarray}
    \vec{\tilde w}_{k}(s_{0})
                     &=&
                       -\left[
             {\bf{D}}(s_{0}+C,\,s_{0})
            -\hat{\lambda}_{k}
                        \right]^{-1}
%Mods for killing splitting
%    \nonumber\\ KILL
%            &\times& {\bf{\tilde{G}}}(s_{0}+C,\,s_{0})\cdot
                      {\bf{\tilde{G}}}(s_{0}+C,\,s_{0})\cdot
%Mods for killing splitting
             \vec{\tilde v}_{k}(s_{0})
\nonumber
\end{eqnarray}
for ($k=I,\ II,\ III$)
and we use a representation of the
         $\bf{\tilde{G}}$ matrix
in the form
\begin{eqnarray*}
        {\bf{\tilde{G}}}(s_{2},\,s_{1})
                 &=&
           \int_{s_{1}}^{s_{2}}{d\tilde{s}}~
                {\bf{      {D}}}(s_{2},\,\tilde{s})\,
                {\bf{\tilde{G}}}_{0}(\tilde{s})\,
                {\bf{\tilde{M}}}(\tilde{s},\,s_{1})
\end{eqnarray*}
with
\begin{eqnarray*}
        {\bf{\tilde{G}}}_{0}
                 &=&
      \left( \begin{array}{rrr}
          l_{s}  & l_{x}  &
          l_{y}               \\
         -m_{s}  & -m_{x} &
         -m_{y}
              \end{array}
       \right)\cdot
       {\bf{\tilde{F}}}
\end{eqnarray*}

In thin lens approximation
the
      $\bf{\tilde{G}}$ matrix
for a quadrupole is
\begin{eqnarray}
      {\bf\tilde{{G}}}
                      &=&
      \left( \begin{array}{llllll}
          -\tilde{q}l_{y}  & 0 & -\tilde{q}l_{x} & 0 & 0 & \kappa_1  \\
          +\tilde{q}m_{y}  & 0 & +\tilde{q}m_{x} & 0 & 0 & \kappa_2  
              \end{array}
       \right) 
\ \ \ \ \ \
\nonumber
\end{eqnarray}
where $\kappa_1 =  -\tilde{q}l_{y}\eta_1  -\tilde{q}l_{x}\eta_3$
and $\kappa_2 =  +\tilde{q}m_{y}\eta_1  +\tilde{q}m_{x}\eta_3$.
We see that as a result of separating the transverse
coordinates into betatron and dispersion contributions, columns six of 
     $\bf{\tilde{F}}$ and $\bf{\tilde{G}}$
contain terms depending on dispersions.

The strong spin matching condition $\vec{{\tilde w}}_{k}=0$ for suppressing 
depolarization now amounts to setting the
${\bf {\tilde{G}}}(s_{\mu_{l+1}},\,s_{\mu_{l}})\,
\vec{\tilde{v}}_{k}(s_{\mu_{l}})$ to
zero in analogy with Eq.(12).
Then in the  special case, or approximation, of a fully uncoupled optic  and 
by
taking into
account only the depolarizing influence of quadrupoles this is equivalent to
requiring 
\cite{chaoyok81,bs85}:

For horizontal motion:
\begin{eqnarray}
%& &
%     c_{x\tilde{p}}^{(\mp)}
%                \ =\
    &&     -\,
            \frac{
               (1+a\gamma_0)}
                 {\sqrt{2}}
%\nonumber    \\
%\nonumber    \\
%& &
%\hspace*{0.5cm}
    \frac{1}{C}
           \int_{s_{\mu_{l}}}^{s_{\mu_{l+1}}}{d\tilde{s}}
                \sqrt{\beta_{x}(\tilde{s})}
                            g(\tilde{s})
                        e^{ -i\, \psi_{x}(\tilde{s})}
%Mods for killing splitting
%          \nonumber\\ && KILL
% \times \left[\,
         \left[\,
%Mods for killing splitting
      l_{y}(\tilde{s})
     \pm i\,m_{y}(\tilde{s})\,
                         \right]
      e^{\mp i
                \psi_{\rm spin}(\tilde{s})
                                } \ =\ 0
%Mods for killing splitting
\nonumber \\
\end{eqnarray}

For vertical motion:
\begin{eqnarray}
%& &
%     c_{y\tilde{p}}^{(\mp)}
%                \ =\
      &&    -\,
             \frac{
                (1+a\gamma_0)}
                  {\sqrt{2}}
%\nonumber    \\
%\nonumber    \\
%& &
%\hspace*{0.5cm}
    \frac{1}{C}
\int_{s_{\mu_{l}}}^{s_{\mu_{l+1}}}{d\tilde{s}}
%           \int_{0}^{C}{d\tilde{s}}
                \sqrt{\beta_{y}(\tilde{s})}
                            g(\tilde{s})
                        e^{ -i\, \psi_{y}(\tilde{s})}
%Mods for killing splitting
%\nonumber\\ && KILL
% \times \left[\,
         \left[\,
%Mods for killing splitting
      l_{x}(\tilde{s})
     \pm i\,m_{x}(\tilde{s})\,
                         \right]
      e^{\mp i
                \psi_{\rm spin}(\tilde{s})
                                } \ =\ 0
%Mods for killing splitting
\nonumber \\
\end{eqnarray}

For longitudinal motion:
\begin{eqnarray}
%& &
%     c_{y\tilde{p}}^{(\mp)}
%                \ =\
       && -\,
            \frac{
               (1+a\gamma_0)}
                 {\sqrt{2}}
   \frac{1}{C}\,
\int_{s_{\mu_{l}}}^{s_{\mu_{l+1}}}{d\tilde{s}}
%           \int_{0}^{C}{d\tilde{s}}
           \frac{
                 \left[ \alpha_{z}(\tilde{s})+i\right]
                                        }
                {\sqrt{
                      \beta_{z}(\tilde{s})}
                            }
                            g(\tilde{s})
                        e^{-i \psi_{z}(\tilde{s})}
\nonumber \\ 
& &  \times
   \{
    \eta_{y}
    \left[
      l_{x}(\tilde{s})
     \pm i m_{x}(\tilde{s})
                         \right]
%Mods for killing splitting
%\nonumber    \\ KILL
%& & +
     +
%Mods for killing splitting
    \eta_{x}
    \left[
      l_{y}(\tilde{s})
     \pm i m_{y}(\tilde{s})
                         \right]
                 \}
      e^{\mp i  \psi_{\rm spin}(\tilde{s})
                                }
         =0
\ \ \ \ 
\end{eqnarray}

Since in practice  synchrotron motion is well approximated by simple
harmonic motion \cite{bmrw86},
       $\beta_{z}(s)$
is almost independent of $s$ and
       $\alpha_{z}(s) \approx 0$. Then
Eq.(18) may be approximated by
\begin{eqnarray}
%     c_{y\tilde{p}}^{(\mp)}
%              & =&
    &&       -\,
            \frac{
               (1+a\gamma_0)}
                 {\sqrt{2}}
           \frac{i}
                {\sqrt{
                      \beta_{z}}
                            }
   \frac{1}{C}\,
\int_{s_{\mu_{l}}}^{s_{\mu_{l+1}}}{d\tilde{s}}
%           \int_{0}^{C}{d\tilde{s}}
                            g(\tilde{s})e^{ -i \psi_{z}(\tilde{s})}
\nonumber\\
\nonumber\\
      &\times& e^{\mp i \psi_{\rm spin}(\tilde{s})}  
 \left\{
    \eta_{y}    \left[
      l_{x}(\tilde{s})
     \pm i\,m_{x}(\tilde{s})
                         \right]
%Mods for killing splitting
%\right.
%\nonumber   \\ KILL
%& &
%\left.\ \ \ \
%Mods for killing splitting
           +
    \eta_{x}
    \left[
      l_{y}(\tilde{s})
     \pm i\,m_{y}(\tilde{s})
                         \right]
                 \right\} \ =\ 0
%\nonumber   \\
%         &=& 0 
\end{eqnarray}

Harmonic synchrobeta spin matching in terms of beta functions and dispersion
follows the path detailed earlier under 
``Alternative Stage 1''   but with the 
eigenvectors $\vec{\tilde v}_{k}$ and the matrices $\bf\tilde F$.
 Typical expressions
can be found in \cite{bs85,yok82}.
\\
\\
\noindent
{\it Commentary} 
\\
\\
Spin matching should be carried out using thick lenses so that the optic is
correct.
Strong spin matching by minimizing the integrals in Eqs.(16--18)
requires explicit integration.
Furthermore Eqs.(16--18) must be modified if there is significant orbital
coupling. Thus
in practice the numerical fitting involved in strong spin matching can be
carried
out most simply  by minimizing the 
%  ${\bf{\tilde{{G_{ij}}}}}(s_{\mu(l+1)},\, s_{\mu l})$ or the 
 ${{{{G_{ij}}}}}(s_{\mu_{l+1}},\, s_{\mu_{l}})$ since these already 
represent integrals and do not need knowledge of the dispersion. Moreover
these matrices are precisely those contained in the SLIM program so that
cross checks between programs are simplified. Another advantage of working 
with the $\bf G$ matrix is that it allows sections of the ring to be 
studied and made transparent in isolation since no knowledge of
Courant--Snyder
parameters is needed; use of $\bf G$ emphasizes the local nature of spin 
transparency. On the other hand  Eqs.(16--18) and the split--up
versions depend on Courant--Snyder parameters and these in turn depend
on the
structure of the whole ring so that the {\it ``locality''} is masked.
When studying the spin transparency of a ring, it is often useful 
for diagnostic purposes to
set elements of the $\bf G$ or the $\bf\tilde{G}$  matrices 
to zero artificially and thereby 
obtain an impression of which sections of the ring are most dangerous.
For example by switching off column six of $\bf\tilde{G}$ in quadrupoles, the
effect of dispersion can be cleanly separated  from the effect of betatron 
motion. One can also investigate the system by using the matrix handling
facilities in symbolic algebra programs and the fact that the 
$\bf G$ and $\bf\tilde{G}$ of magnets or strings of magnets often depend
in a simple way on the elements of the  corresponding 
$\bf M$ and $\bf\tilde{M}$ \cite{bar85b}.
Finally, the $\bf G$ and  $\bf\tilde{G}$  matrices are in general energy 
dependent.
But a spin match made at the design energy is usually still effective 
for a few tens of MeV above and below, except near resonances. 
\\
\\
%\newpage
\noindent
{\it Some examples}
\\
\\
In a perfectly aligned flat ring (no vertical bends) with no solenoids and 
no $x-y$ coupling, the depolarization rate $\tau_{\rm dep,lin}^{-1}$ 
vanishes (see below under {\it Harmonic closed orbit spin matching}) 
so that no spin matching is needed.

A spin rotator (Secs.2.7.3, 2.7.4 in \cite{handbookb})
 based on dipoles and containing no
 quadrupoles
is automatically almost spin transparent since the 
elements of $\bf G$ are usually much  smaller in dipoles than in 
quadrupoles \cite{bar85b}. 
Dipole rotators containing quadrupoles need
explicit spin matching \cite{grote94}.

Spin rotators based on a combination of solenoids (which rotate $\hat{n}_{0}$
from the vertical into the horizontal) and dipoles (to make the polarization
longitudinal at an interaction point (IP)) \cite{bar85b} are not 
automatically transparent. They also cause $x-y$ coupling. 
However,
by sandwiching quadrupoles and skew quadrupoles among sections of solenoid
the coupling can be eliminated and by careful choice of the sandwich 
structure some terms in columns 1 to 4 
of $\bf G$  for the rotator can be made small at the same time 
\cite{bar85b}.
Column 6 remains troublesome  but for antisymmetric solenoid schemes 
\cite{bar85b} the columns 6 of the rotators  cancel each other. 
For further discussion on solenoids see \cite{zh81,bar822}. 

For a straight section (e.g. surrounding an IP) where the 
polarization is longitudinal and which only contains quadrupoles and drifts,
the spin precession angle is
a linear combination of the overall orbit deflections 
${\Delta p}_x$ and ${\Delta p}_y$ in the quadrupole fields \cite{bar85b}.
Thus spin transparency implies making ${\Delta p}_x$ and ${\Delta p}_y$
vanish for all orbits. 
This can also be deduced from Eqs. (16) and (17). 
If the straight section is geometrically and 
optically left--right symmetric, this can be achieved with an optic for which
$\tan{{\Delta \psi}_x}=-{\alpha}_x$ and $\tan{{\Delta \psi}_y}=-{\alpha}_y$
 where
the $\Delta \psi$ are the phase advances between the IP and an
outer end of the straight section and the $\alpha$ appertain to the
outer end.  So the eight conditions that columns 1 to 4 of the 
$\bf G$ matrix vanish have been reduced to two conditions by the symmetry.
Furthermore, this is an example where the spin matching conditions reduce
to purely optical conditions.

These conditions can also be formulated directly in terms of $\bf G$. By
choosing ${\hat{l}} = \hat y$ and ${\hat{m}} = \hat x$
and requiring that 
the elements $G_{11}$ and $G_{23}$ vanish for the stretch  from the IP
to the outer end, $\bf G$ vanishes for the whole straight 
section for an arbitrary orientation of ${\hat{m}}$, ${\hat{l}}$ around the
longitudinal ${\hat{n}}_0$. 

For a straight section modified to contain horizontally
bending dipoles with ${\hat{n}}_0$ in the horizontal plane, Eq. (19)
is equivalent to requiring that the total change of $\eta_2$ due to the
quadrupoles vanishes  over the section.

If the straight section contains rf cavities, their influence
on the spin transparency can often be neglected. 

Other examples of the use of symmetry to simplify the spin match can be 
found in \cite{bs85} where spin matching using variants of Eqs.(16--18) for a
ring with
dipole rotators is discussed. The results of a calculation with SLICK before
and after a spin match can be found in \cite{bonn90}. Experimental 
observations resulting from successful spin matching involving spin
rotators are described in \cite{bar95b}.
\\
\\
\noindent
{\it Computer programs for strong spin matching} 
\\
\\
Strong spin matching facilities  based on evaluation of spin--orbit integrals
(e.g. Eqs.(16--18) ) are built into the programs ASPIRRIN and 
SOM. To do spin matching in terms of $\bf G$ the code SPINOR 
\cite{hs} can be used.
\\
\\
\noindent
Stage 2: {\it Harmonic closed orbit spin matching} 
\\
\\
Once the perfectly aligned ring has been spin matched, the effects of
misalignment must be addressed. 
In a perfectly aligned flat ring with no solenoids,
$\hat{n}_{0}$ is vertical so that  $l_{y}$ and  $m_{y}$
are zero. Then by inspection of the $\bf G$ matrix elements for 
horizontal bends, quadrupoles and rf cavities it is clear 
that for no $x-y$ coupling,  columns 1, 2, 5 and 6 of ${\bf{G}}(s+C,\,s)$
vanish. In particular, for quadrupoles, columns 1 and 2 of ${\bf{G}}$ and columns 1, 2 and 6 of
$\bf{\tilde{G}}$ vanish.
Moreover with no $x-y$ coupling the one turn orbital matrix 
$\bf{M}_{6\times 6}$  and its eigenvectors have the structures \cite{mais82}  
\begin{eqnarray}
   \bf{M}_{6\times 6}
                  &=&
                 \left( \begin{array}{cccccc}
\star  &  \star  &  0     &  0      &  \star  &  \star  \\
\star  &  \star  &  0     &  0      &  \star  &  \star  \\
0      &  0      & \star  &  \star  &   0     &  0      \\
0      &  0      & \star  &  \star  &   0     &  0      \\
\star  &  \star  &  0     &  0      &  \star  &  \star  \\
\star  &  \star  &  0     &  0      &  \star  &  \star
              \end{array} \right)\ ;
%\nonumber
%\end{eqnarray}
%\begin{eqnarray}
             ~~\vec{v}_{I}
                =
       \left( \begin{array}{c}
                \star \\
                \star \\
                0     \\
                0     \\
                \star \\
                \star
                \end{array}     
         \right)    \, ;
            ~~\vec{v}_{II}
                =
       \left( \begin{array}{c}
                0    \\
                0     \\
                \star \\
                \star \\
                0     \\
                0    
                \end{array}     
         \right)    \, ;
           ~~\vec{v}_{III}
                =
       \left( \begin{array}{c}
                \star \\
                \star \\
                0     \\
                0     \\
                \star \\
                \star
                \end{array}
         \right)
\nonumber
\end{eqnarray}
where a $\star$ denotes a nonzero element.
Therefore by Eq.(6) $\vec{w}_{I}(s)$ and $\vec{w}_{III}(s)$ are zero. 
Note that  for no $x-y$ coupling $v_{II5}^{*} \equiv v_{y5}^{*}$
vanishes. Then by Eq.(10)
$\tau_{\rm dep,lin}^{-1}$
is  automatically zero. In rings with vertical bends (e.g. in spin rotators) 
$\hat{n}_{0}$ is made vertical in the arcs by design.

In real misaligned rings there is a vertical closed orbit
distortion and 
         $\hat{n}_{0}$
is tilted from the vertical in the arcs (see below) 
so that the above mentioned  columns of
         $\bf{G}$
and 
         $\bf{\tilde{G}}$
for the arc quadrupoles do not vanish. 
In practice the tilts can be tens of milliradians and they increase
with energy (they are roughly proportional to $a\gamma_0$)
but even these small angles
can lead to strong depolarization so that  it is essential that
the ring be very well aligned from the beginning.
Note that {\it vertical} closed orbit distortion leads 
primarily to depolarization
due to {\it horizontal} synchrobetatron motion in the arcs.
Note also that tilts of tens
of milliradians cause a negligible decrease of the underlying ST 
polarization (Eq.(14), Sec.2.7.7).

If there is a vertical correction coil and a beam position monitor (BPM)
near each quadrupole, one can try to minimize the combined vertical kick 
({\it ``kick minimization''}) \cite{epac96} applied to the orbit 
by each quadrupole and its correction coil
and thereby reduce the tilt of $\hat{n}_{0}$ due to the
distorted orbit's being off centre in the (misaligned) quadrupoles. 
This also reduces the generation of spurious vertical dispersion so that the
driving of $\nu_y$ and $\nu_z$ resonances (Eq.(11)) is avoided.
This presupposes that the positions with respect to 
the quadrupoles of the BPMs are well known. These relative 
positions can be estimated using beam--based calibration
(Sec.4.5.5 in \cite{handbookb})\cite{epac96}. However, kick minimization 
will not be effective if, say, the  dipoles have significant tilt 
misalignments.

If these measures are insufficient, a further method for bringing
         $\hat{n}_{0}$
closer to the vertical is needed. 
         $\hat{n}_{0}$, and thus its tilt, for the distorted ring
can be obtained as described in Sec.2.7.7  but
 one gains more insight by using
 a perturbation theory based on SLIM concepts \cite{bmrr}.
Viewed from the  $(\hat{n}_{0},\, \hat{m},\, \hat{l})$ frame calculated for
the design orbit, the first order
 deviation of
         $\hat{n}_{0}$
from the design orientation can be written as
\begin{eqnarray}
      \left[
             \delta n_{01}(s)-i\delta n_{02}(s)
                  \right]
                &=&
           -i\frac{C}{2\pi}
  \sum_{k}
        h_{k}
       \frac{ e^{
         i{2}\pi k {s}/{C}  }    }
            {k-\nu_{\rm spin}}
\nonumber
\end{eqnarray}
where the
         $h_{k}$
are Fourier coefficients given by
\begin{eqnarray}
         h_{k}
                 &=&
        \frac{1}{C}
           \int_{s_{0}}^{s_{0}+C}d\tilde{s}
      \left[
            d_{1}(\tilde{s})
                   -
            i d_{2}(\tilde{s})
                \right]
                        e^{
          -i{k}2\pi \tilde{s}/{C}
                                                   }
\nonumber
\end{eqnarray}
Here
\begin{eqnarray}
&   
    \left( \begin{array}{c}
              d_{1}     \\ d_{2}
                \end{array}
         \right)
             \ =\
                        \left( \begin{array}{rrr}
                           l_{s}  & l_{x}  &  l_{y}      \\
                          -m_{s}  &-m_{x}  & -m_{y}
              \end{array}
         \right)
%Mods for killing splitting
%\nonumber\\ KILL
% &\cdot\
%Mods for killing splitting
      \left\{
         {\bf{F}}\cdot\vec{u}_{{\rm co}}
              -
        \frac{e}{p_{0}}
       \left( \begin{array}{c}
               \Delta B_{s}\frac{1+a\gamma_0}{1+\gamma_0} \\
               \Delta B_{x}(1+a\gamma_0) \\
               \Delta B_{y}(1+a\gamma_0)
                \end{array}
         \right)
              \right\}
\ \ \ \ \
\nonumber
\end{eqnarray}
where the $\Delta B_{x,y,s}$ are field errors and $\vec{u}_{{\rm co}}$
is the deviation of the 6--D closed orbit from the design orbit.
     $\delta{\hat{n}}_{0}$
can be minimized by using correction coils to adjust the closed orbit
(e.g. by generating closed bumps so that the luminosity is not affected)
in such a way that the real and imaginary parts of
        $h_{k}$,
with
        $k$
near
        $\nu_{\rm spin}$,
are small.
This technique is called
   {\bf harmonic closed orbit spin matching} and is embodied in the program 
FIDO \cite{mane85,hpol2b}. See \cite{rs85} also. 
If the machine distortions
are not well known and if the closed orbit cannot be measured well enough,
the closed orbit correction must be carried out empirically by observing
the polarization. If the distortions and the orbit are well enough known
the  correction coil strength can be calculated {\it ab initio} 
(deterministic harmonic closed orbit spin matching) \cite{plac94}.
The correction scheme should be chosen so that it achieves the maximum 
effect on  $\delta {\hat{n}}_{0}$ with the smallest possible additional orbit
 distortion.

Harmonic closed orbit spin matching can in principle be used to minimize
the  $\delta{\hat{n}}_{0}$  due to an uncompensated solenoid
placed at the position of a nominally vertical ${\hat{n}}_{0}$. However, this 
is achieved more efficiently by generating relatively antisymmetric 
vertical orbit bumps (spanning horizontal bend magnets) on each  side of
the solenoid \cite{ks82,blond}.  

It might also be useful to weight $\delta\hat{n}_{0}(s)$ 
by a periodic function $p(s)$ \cite{br87}.
In that case one tries to minimize
         $p(s)   \delta\hat{n}_{0}(s)$.
This is worth trying, for example, if the main source of depolarization 
due to misalignments 
is the coupling of non-zero  $l_{y}$ and $m_{y}$ to the horizontal
dispersion in the arcs (see Eq.(19)). This is often the case, as can be seen 
by examining the numerical values of the contributions of each
mode ($I, II, III$) in Eq.(10). Then $p(s)$ is taken to be $\eta_x(s) g(s)$.

To minimize $ p(s)\delta\hat{n}_{0}(s)$ one must minimize the harmonics 
${\tilde h}_{k}$  of
\begin{eqnarray}
{\tilde h}(s)
      &=& 
    p(s)(d_{1}-i d_{2})
%Mods for killing splitting
%\nonumber  \\ KILL
%      & &\ \ \ \ \ +\,
+
%Mods for killing splitting
    p'(s)
      \left[
             \delta n_{01}(s)-i \delta n_{02}(s)
                  \right]
\nonumber   \\
\nonumber   \\
     &=&
   {\tilde h}(s+C)
\nonumber
\end{eqnarray}
whereby
\begin{eqnarray}
& 
   p(s) \left[ \delta n_{01}(s)-i\delta n_{02}(s) \right]
%Mods for killing splitting
%\nonumber   \\ KILL
%&=\
\ =\
%Mods for killing splitting
           -i\frac{C}{2\pi}
  \sum_{k}\,
 {\tilde h}_{k}
       \frac{ e^{i{2}\pi k{s}/{C}  }   }  
            {k-\nu_{\rm spin}} 
\nonumber
\end{eqnarray}
\\
\\
\noindent
Stage 3: {\it Further tuning} 
\\
\\
Harmonic closed orbit spin matching can generate spurious vertical dispersion
and this in turn generates vertical emittance 
(nonzero  $v_{II5}^{*}$ (Sec.2.1.4 in \cite{handbookb})) 
and also ensures that column 6 of $\bf{\tilde{G}}$ for the quadrupoles does not
vanish. Thus 
extra depolarization can occur. It might then be useful to overlay
a harmonic vertical betatron match ($k = II$ in Eq.(15)) 
on any existing
Stage 1 match, assuming that is possible.
Likewise, to overcome the effect of spurious vertical dispersion in
column 6 of $\bf{\tilde{G}}$ 
 one could use extra vertical correction coils to overlay a  
harmonic vertical dispersion match ($k = III$ in Eq.(15)). Usually both
of these two extra matches would be empirical. One could also try to 
combine the harmonic closed orbit match and  the harmonic vertical dispersion
match into one procedure.
\\
\\
\noindent
Stage 4: {\it Beam--beam spin matching} 
 \\
\\
The beam--beam interaction is equivalent to a nonlinear lens and can  
spoil a spin match.  The effect of the beam--beam interaction on  the 
polarization is not fully understood but it has been suggested that the 
beam--beam depolarization can be reduced by
balancing the beam--beam deflection of  spins against 
subsequent deflections taking place
in the ring quadrupoles. The condition for minimizing the effect of 
vertical kicks is independent of the current and charge distribution 
in the opposing beam and takes the form \cite{buon84}
\begin{eqnarray}
& &
\frac{m_{x}-il_{x}}{\sqrt{\beta_{y}^{*}}}+
\sum_{\pm}\,\pm\frac
{e^{-\frac{i}{2}(\nu_{\rm spin}\pm\nu_{y})}}
{4 \sin\frac{\nu_{\rm spin}\pm\nu_{y}}{2}}
%Mods for killing splitting
%\nonumber KILL
%\\
%&\times&   
%Mods for killing splitting
  \int_{0}^{C}{ds}~g~\sqrt{\beta_{y}}
                            e^{\pm i\psi_{y}}
                           ~(m_{x}+il_{x}) \ =\ 0
%\nonumber
%\\        
%       &=&0  \ .
\nonumber
\end{eqnarray}
An equivalent prescription in SLIM formalism allows an arbitrarily coupled 
optic to be treated \cite{bar952}. 

\vspace{5mm}
\paragraph{Higher order resonances}
To go beyond the linearization of spin contained in Eq.(1) one writes
\begin{equation}
            \hat{n}(\vec{\tilde u}; s)
                 \ =\
            (1 - {\alpha}^2 -{\beta}^2)^{1/2} \hat{n}_{0}(s)
                  +
            \alpha \hat{m}(s)
                  +
            \beta  \hat{l}(s)
\end{equation}
(for $\alpha^2 + \beta^2 \leq 1 $) 
and does not linearize the T--BMT equation. Then spin--orbit resonances
of arbitrarily high order can appear in 
 $\frac{\partial{\hat{n}}} {\partial{\delta}}$ \cite{mane87b}. 
The strength decreases with the order
 ($\equiv |k_{I}| + |k_{II}| + |k_{III}|)$. 
In practice the most intrusive higher order resonances are those for which 
$ \nu_{\rm spin} = k_0 \pm \nu_{k} + k_{III}\nu_{III}$. These
{\it ``synchrotron sideband resonances''} of the first order parent resonances
are due to modulation by energy oscillations of the instantaneous rate of 
spin precession around 
$\hat{n}_{0}$. They  originate in
the part due to synchrotron motion in the 
term $\vec{\omega}^{\rm sb}\cdot\hat{n}_{0}$ appearing in the full equations
of spin motion (i.e. beyond the SLIM level) \cite{yok83}. 
The depolarization rate 
associated with sidebands of isolated parent resonances
($ \nu_{\rm spin} = k_0 \pm \nu_{k}$)
is approximately proportional to the depolarization rate for the parent 
resonances. 
Thus  the effects of 
synchrotron sideband resonances can be reduced by doing the spin matches
described above. Explicit formulae for the proportionality constants  
({\it ``enhancement factors''})  can be found in  
\cite{mane90,mane92}. The underlying strength parameter 
(the {\it ``modulation index''}) of synchrotron sideband
resonances is $(a\gamma_0 {\sigma}_{\delta}/{\nu}_z)^2$ which increases 
strongly with the energy and  energy spread.

\vspace{5mm}
\paragraph{Other computer codes \protect{\cite{hvb99b}}}
The SMILE algorithm is restricted to linearized orbital motion in the thin lens approximation and
calculates $\frac{\partial{\hat{n}}} {\partial{\delta}}$ 
by an extension of the first order perturbation theory of SLIM to
high order using Eq.(20) and full 3--D spin motion.
The algorithm involves multi--turn spin--orbit tracking.  High
order resonance effects are manifested by resonance denominators but the 
formalism ensures that the vector $\hat n$ is of unit length.
The highest required absolute values of the $k_{I}, k_{II}, k_{III}$ 
are specified as  input parameters. 

SODOM represents $\hat n$ by a spinor notation. The periodicity
condition $\hat{n}(\vec{u}; s)  =  \hat{n}(\vec{u}; s+C)$ (Sec.2.7.7) 
is equivalent to periodicity in the three phases of linearized orbital
motion and the one turn 2 $\times$ 2 spinor transfer matrix on a synchrobeta
 orbit is
also periodic in the initial orbital phases. The spinor transfer matrix and
$\hat{n}(\vec{u}; s)$ are then represented by Fourier series. 
The Fourier coefficients are obtained numerically and 
$\hat{n}(\vec{u}; s)$ can then be reconstructed. By constructing  $\hat n$
at many points in phase space 
$\frac{\partial{\hat{n}}} {\partial{\delta}}$ can be obtained by numerical 
differentiation.The highest required absolute values of the
$k_{I}, k_{II}, k_{III}$ are specified as input parameters. 

The algorithm SpinLie utilizes Lie algebraic methods 
(Sec.2.7.9 in \cite{handbookb}) 
to provide a perturbation expansion for $\hat n$ and can handle 3-D spin motion and
moderately non-linear orbit motion.

The vector $\hat{n}(\vec{u}; s)$ can also be obtained by 
{\it ``stroboscopic averaging''}~using the code SPRINT. 
$\frac{\partial{\hat{n}}} {\partial{\delta}}$ can then be calculated by 
numerical differentiation. This algorithm automatically includes all
orders of resonance.

The above algorithms all exploit the DKM 
formula (Eq.(16),Sec.2.7.7) but
the SITROS   and SLICKTRACK algorithms simulate the depolarization process directly using
Monte--Carlo tracking simulations of the effects on the orbit, and then on the
spin, of
stochastic photon emission and damping and deliver estimates of
$\tau_{\rm dep}$. The equilibrium polarization is then obtained from the  
approximation (Sec.2.7.7)
\begin{eqnarray}
      P_{\rm eq} &=& P_{\rm bks} \frac{\tau_{\rm tot}}{\tau_{\rm bks}}
\end{eqnarray}
where 
\begin{eqnarray}
      \frac{1}{{\tau_{\rm tot}}}
           &=&
      \frac{1}{\tau_{\rm bks}}
            +
      \frac{1}{\tau_{\rm dep}}\ .
\end{eqnarray}
This ignores the (normally small) term
 $\frac{\partial{\hat{n}}} {\partial{\delta}}$
in the numerator of the DKM formula.
SITROS and SLICKTRACK calculate with full 3-D spin motion and, 
in contrast to the analytical algorithms, they can handle strongly nonlinear orbital motion.


\begin{thebibliography}{42}
\bibitem{bar85b}
D.P. Barber  et al., Particle Accelerators  17 (1985) 243.
Note that in this paper, the symbol $\vec n$ should be
replaced with $\vec n_0$ to correspond with modern notation.
\bibitem{mais82}
We are indebted to H. Mais for discussions on this point.
\bibitem{bmrr}
D.P. Barber  et al., DESY Report 85-44 (1985).
Note that in this paper, the symbol $\vec n$ should be
replaced with $\vec n_0$ to correspond with modern notation.
\end{thebibliography}
\end{document}